\documentclass[sigconf]{acmart}

\copyrightyear{2026}
\acmYear{2026}
\setcopyright{cc}
\setcctype{by}

\acmConference[ICAIF '26]
  {7th ACM International Conference on AI in Finance}
  {November 14--17, 2026}
  {Milan, Italy}
\acmBooktitle{7th ACM International Conference on AI in Finance
  (ICAIF '26), November 14--17, 2026, Milan, Italy}
\acmDOI{10.1145/3858063.3858699}
\acmISBN{979-8-4007-3087-0/2026/11}

\usepackage{amsmath,mathtools}
\usepackage{booktabs,graphicx,tabularx}
\newcolumntype{Y}{>{\raggedright\arraybackslash}X}

\newtheorem{theorem}{Theorem}
\newtheorem{definition}{Definition}

\theoremstyle{remark}
\newtheorem{remark}{Remark}

\newcommand{\R}{\mathbb{R}}
\newcommand{\A}{\mathcal{A}}
\newcommand{\eps}{\varepsilon}
\newcommand{\Risk}{\mathcal{R}}
\newcommand{\NNProj}{\ensuremath{\mathrm{NN\text{-}Proj}}}
\newcommand{\st}{\;:\;}
\AtBeginDocument{}

\begin{document}

\title{Quantifying Repair Risk in Crypto Option Surfaces} 

\author{Tenghan Zhong}
\affiliation{%
  \institution{University of Southern California}
  \city{Los Angeles}
  \state{California}
  \country{USA}
}
\email{tenghanz@usc.edu}

\begin{abstract}
Option desks routinely adjust raw crypto option quotes into price surfaces that
remain within bid--ask intervals and satisfy standard no-arbitrage restrictions
across strikes before computing Greeks, hedge quantities, or risk measures.
For the same quote panel, however, these constraints may admit multiple such
surfaces, which we call \emph{admissible repairs}, with different downstream
outputs. We define the resulting variation as \emph{repair risk} and propose \emph{repair-risk validation}, a matched framework for quantifying dispersion in Greeks, hedge quantities, and risk measures.

The mechanism is finite-difference amplification: two admissible repairs within
price distance \(\eps\) can differ in local normalized-strike curvature by
\(4\eps/h^2\), where \(h\) is local normalized-strike spacing.
Motivated by this mechanism, we evaluate repair risk on Deribit BTC and ETH
option chains, using \(L_1\) and
\(L_2\) projections of midpoint prices onto the same admissible set as matched
controls. Across 108 matched evaluations spanning 18 sampled calendar months, the
median 90th percentile of the scaled spot-gamma gap for \(L_2\) versus the
\(\rho=10\) smoothness-penalized repair is nearly twenty times the matched
\(L_1\)--\(L_2\) control. We also instantiate \NNProj{}, a neural-projection repair, to test whether
learned repairs exhibit repair risk. Despite good held-out fit,
its downstream outputs
remain distinct from those of classical repairs. Across five seeds with the same training window and architecture, its curvature
dispersion exceeds the matched \(L_1\)--\(L_2\) control in 17 of 18 months
(\(p=1.45\times10^{-4}\)). Economically, for a USD~1 million risk reversal, switching from
\(L_2\) to the \(\rho=10\) repair shifts the median 90th-percentile hedge
notional by 18.62 basis points of gross spot notional. VaR-style estimates also
exhibit greater dispersion across repairs. The repair rule is therefore part of the risk model, not merely
a feasibility check.
\end{abstract}

\ccsdesc[500]{Applied computing~Economics}
\ccsdesc[500]{Computing methodologies~Model verification and validation}
\ccsdesc[300]{Computing methodologies~Machine learning}
\ccsdesc[100]{Mathematics of computing~Mathematical optimization}

\keywords{option surfaces, repair risk, no-arbitrage constraints, neural repair, model validation, crypto options, model risk, trustworthy AI}

\interlinepenalty=10000
\maketitle
\interlinepenalty=0

\section{Introduction}

Raw crypto option quotes are rarely used directly in valuation and risk
calculations. They can be wide, sparse, and violate no-arbitrage conditions
across strikes. We therefore convert them to a normalized call-price vector and
repair that vector to remain within bid--ask intervals and satisfy those
conditions. We call any price vector
satisfying these constraints an \emph{admissible repair}, and write the
feasible set as \(\A\).

This step is usually treated as a price-feasibility screen. Most option-surface
repair and arbitrage-free construction methods validate the repaired surface by
price fit, feasibility, or smoothness. Recent neural surface methods are
likewise often validated by fit and feasibility rather than downstream
agreement.
We instead ask a downstream question: after several classical or learned repairs
satisfy the same feasibility constraints, do they agree on the Greeks, hedge
quantities, and risk measures computed from the repaired surface?

We call this dispersion \emph{repair risk}: for any scalar downstream
quantity, it is the range of that quantity across admissible repairs of the
same quote panel. This is an identification problem: a single feasible set can
contain multiple admissible repairs that satisfy the same no-arbitrage
constraints across strikes yet disagree on Greeks, hedges, or risk measures.
Empirically, pairwise differences across the repair rules considered below
provide observed lower bounds on the full feasible-set range. For local
curvature, we compute the full range directly by linear programming.

A key mechanism behind this curvature range is finite-difference amplification. We show that two admissible repairs at sup-norm price distance \(\eps\) can
differ in local curvature with respect to normalized strike \(x=K/F\) by
\(4\eps/h^2\), where \(h\) is the local spacing in \(x\). A repair negligible in price can therefore be material for curvature-based
Greek estimates. Since no-arbitrage constraints across strikes do not uniquely select an
admissible repair, any implemented repair requires an additional selection
criterion.

Empirically, we study the consequences of this nonuniqueness on Deribit BTC
and ETH option surfaces.
The matched design holds the quote panel, feasible set, book templates, and downstream functionals fixed and
varies only the repair rule. \(L_1\) and \(L_2\)
projections serve as stable controls, while smoothness-penalized repairs
separate further from \(L_2\) as the penalty rises. \NNProj{}, a learned repair
obtained from a neural fit to total implied variance followed by projection
into \(\A\), remains
distinct from both \(L_2\) and the \(\rho=10\) smoothness-penalized repair, denoted
\(\mathrm{Smooth}_{10}\), despite good out-of-sample fit in log total implied variance. The theory motivates the ranking score (\(\mathrm{RelSpread}/h^2\)), which
combines relative bid--ask spread with local normalized-strike spacing to rank slices by their potential for curvature amplification. The consequences are economic: the
repair choice shifts hedge notionals on standardized books and produces material dispersion in VaR-style estimates.

This paper makes four contributions. First, we formulate \emph{repair risk}
as an identification problem: multiple admissible repairs of a single quote
panel can satisfy the same constraints yet produce different Greeks,
hedges, and risk measures. Second, a locally sharp construction gives the
\(4\eps/h^2\) curvature separation, and linear programming characterizes
the full feasible curvature range. Third, a separate 18-month replication
using five training seeds shows that good predictive fit and projection
into the same feasible set do not ensure stable outputs from a learned
repair: neural refits still produce materially different admissible
surfaces and curvature. Fourth, we quantify economic magnitude in BTC and
ETH options by measuring how repair choice changes standardized-book
hedges and VaR-style estimates.

\paragraph{Code Availability.}
The code repository is available at
\url{https://github.com/TenghanZhong/no-arb-code}.

\section{Related Work}

\paragraph{Option-surface feasibility and repair.}
Finite-strike no-arbitrage, filtering, and repair methods characterize or
project traded prices into feasible sets
\cite{davis2007range,cohen2020detecting,moussa2026filtering}.
Arbitrage-free interpolation methods add construction choices beyond static
feasibility \cite{kahale2004arbitrage,andreasen2011volatility}, while smoothing
and parametric surface models impose additional regularity or low-dimensional
structure \cite{fengler2009arbitrage,gatheral2014arbitrage}. This literature
treats feasibility, fit, and smoothness as central design goals. Our focus is repair risk after feasibility has already been imposed: among
repairs within the same quote bounds and feasible set, how much can Greeks,
hedges, and risk measures still differ?

\paragraph{Risk-neutral densities and smoothing sensitivity.}
The risk-neutral-density literature starts from the link between strike
derivatives of option prices and state prices \cite{breeden1978prices}.
Nonparametric pricing and shape-restricted estimators show that density recovery
depends on smoothing and no-arbitrage restrictions
\cite{ait_sahalia_lo_1998,aitSahaliaDuarte2003shape}. Empirical implied-density
work further shows that recovered distributions and moments can move with
estimator choice and quote perturbations
\cite{bliss_panigirtzoglou_2002}. We recast this smoothing sensitivity as repair risk among admissible repairs
within a fixed feasible set.

\paragraph{Crypto options and model uncertainty.}
Deribit inverse options and USD/coin accounting make repair and hedge measurement especially relevant in crypto option markets
\cite{lucicSepp2024inverse}. We view repair risk as a derivative-pricing form
of model uncertainty \cite{cont2006model}: the modeling choice is the rule that
selects one admissible repaired surface from a fixed quote panel and feasible
set.

\paragraph{Model multiplicity and learned surfaces.}
In machine-learning terms, the closest framing is underspecification and
predictive multiplicity
\cite{damour2022underspecification,marx2020predictive}: multiple predictors can
pass similar validation criteria yet differ on consequential outputs. We apply
this lens to option-surface repair rules: feasibility is shared, while
curvature, hedges, and risk measures may differ. Machine-learning methods have
been used to learn option and implied-volatility surfaces
\cite{ackerer2020deep,chataigner2021shape,ning2023arbitrage}. Related work also
includes physics-informed calibration of intraday implied-volatility surfaces
under no-arbitrage constraints \cite{hoshisashi2024wamol}, generative and
operator-based surface constructions
\cite{ginartCole2025gan,wiedemannJacquierGonon2025operator}, and neural term
structures as low-dimensional pricing representations
\cite{linLiu2024neuralTermStructure}. Here, \NNProj{} tests whether repair risk
remains after projecting a learned target into the same feasible set.

\paragraph{Financial AI validation.}
The paper also relates to model-risk and trustworthy-AI frameworks in finance
\cite{nist2023airmf}. These frameworks emphasize that
predictive fit alone is insufficient when modeling choices affect operational
decisions. We extend this principle to learned option-surface repair: good fit
and no-arbitrage feasibility need not ensure stable Greeks, hedge quantities,
or risk measures.

\section{Repair Risk and Feasible-Set Mechanism}

This section isolates the local mechanism on a uniform normalized-strike grid;
the empirical analysis uses the corresponding nonuniform three-point stencil.
We proceed from feasibility to curvature amplification, its implications for
Greek outputs, and bounds on local curvature repair risk.

\subsection{Single-Maturity Feasible Set}

Fix one maturity. Let \(N\) be the number of strike intervals, \(h>0\) the
uniform normalized-strike grid spacing, \(x_0\) the leftmost normalized strike,
\(x_j=x_0+jh\), \(j=0,\ldots,N\), and
\(c=(c_0,\ldots,c_N)\in\R^{N+1}\) the normalized call prices. Define
\((D_1c)_j=(c_j-c_{j+1})/h\), \(j=0,\ldots,N-1\), and
\((D_2c)_j=(c_{j-1}-2c_j+c_{j+1})/h^2\), \(j=1,\ldots,N-1\).
Under our sign convention, the normalized price of the adjacent call spread is
\(c_j-c_{j+1}=h(D_1c)_j\), so no arbitrage requires
\(0\le(D_1c)_j\le1\).
The corresponding butterfly-spread price is
\(c_{j-1}-2c_j+c_{j+1}=h^2(D_2c)_j\); hence
\((D_2c)_j\ge0\) enforces discrete convexity.

Let \(\mathrm{bid}_j\) and \(\mathrm{ask}_j\) denote the normalized bid and
ask quotes, and let \(b_j:=\max\{\mathrm{bid}_j,\max\{1-x_j,0\}\}\) and
\(a_j:=\min\{\mathrm{ask}_j,1\}\) denote the effective price bounds.
The single-maturity feasible set is
\[
\A=\{c\in\R^{N+1}\st b_j\le c_j\le a_j,\ 
0\le(D_1c)_j\le1,\ (D_2c)_j\ge0\}.
\]
Each inequality is imposed over its natural index range. We apply
\(\A\) separately at each maturity and suppress the maturity index.
Admissibility here isolates nonuniqueness among repairs satisfying the same
no-arbitrage constraints across strikes; it does not impose cross-maturity
no-arbitrage constraints. A vector is globally strictly feasible when
\(b_j<c_j<a_j\), \(0<(D_1c)_j<1\), and \((D_2c)_j>0\).
Whenever nonempty, \(\A\) is a compact convex polytope; the repair problems are
convex programs, and each local \(D_2\) range can be obtained from two linear
programs.

\subsection{Sharp Curvature Amplification}

For \(\eps\ge0\) and \(c,\tilde c\in\R^{N+1}\) with
\(\|c-\tilde c\|_\infty\le\eps\), the triangle inequality gives, for every
interior \(i\),
\begin{equation}
\label{eq:universal-d2-bound}
\left|(D_2c)_i-(D_2\tilde c)_i\right|\le4\eps/h^2.
\end{equation}
Theorem~\ref{thm:sharp-admissible-amplification} shows that this bound is
locally sharp within \(\A\) itself: if \(c^0\) is globally strictly
feasible, the attaining repairs \(c^+\) and \(c^-\) remain globally
strictly feasible for all sufficiently small \(\eps\).

\begin{figure*}[!t]
\centering
\includegraphics[width=0.78\textwidth]{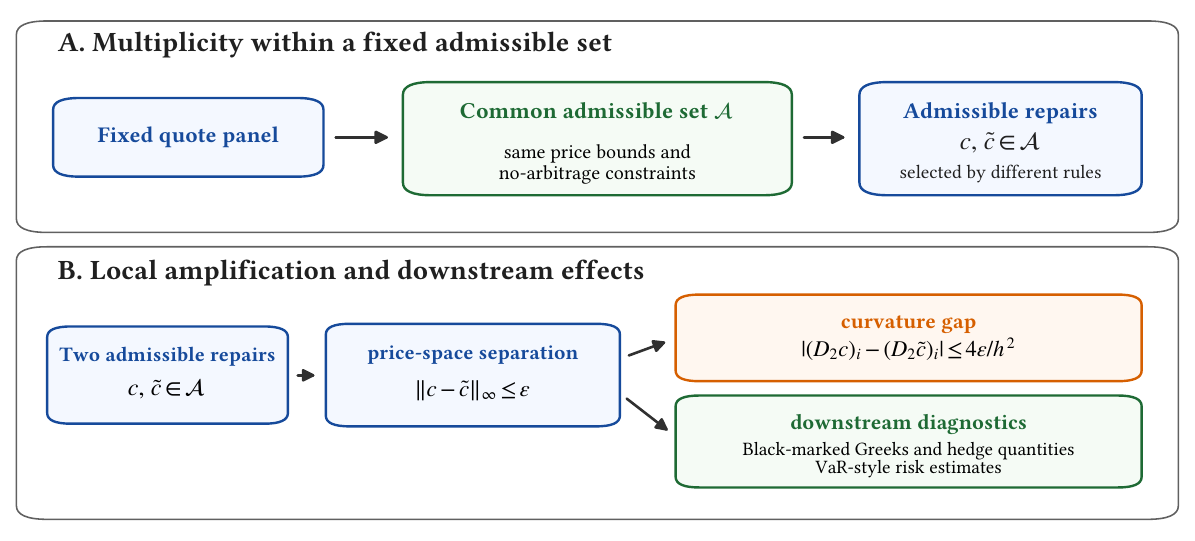}
\Description{Diagram showing multiple repairs in a fixed admissible set, the locally sharp curvature-separation bound, and separate downstream evaluations.}
\caption{Fixed-set multiplicity and local amplification. A fixed quote panel
can admit multiple repairs in the same admissible set; normalized-price
distance \(\eps\) can produce local curvature separation up to
\(4\eps/h^2\). Greeks, hedge quantities, and VaR-style estimates are evaluated
separately for each repair.}
\label{fig:mechanism-overview}
\end{figure*}

\begin{theorem}[Sharp admissible curvature amplification]
\label{thm:sharp-admissible-amplification}
Assume \(N\ge6\) and fix \(i_\star\in\{3,\ldots,N-3\}\). Let \(I_0=\{i_\star-1,i_\star,i_\star+1\}\), \(I_1=\{i_\star-2,\ldots,i_\star+1\}\), and \(I_2=\{i_\star-2,\ldots,i_\star+2\}\) be the affected price, first-difference, and second-difference stencils.
These sets are bookkeeping: the three-point price perturbation changes three prices, four adjacent first differences, and five adjacent second differences.
Suppose \(c^0\in\A\) has local slack \(\kappa,s_1,s_2>0\):
\[
\begin{aligned}
 b_j+\kappa&\le c^0_j\le a_j-\kappa &&(j\in I_0),\\
 s_1&\le(D_1c^0)_j\le1-s_1 &&(j\in I_1),\\
 (D_2c^0)_j&\ge s_2 &&(j\in I_2).
\end{aligned}
\]
Let \(\eta:=\min\{\kappa,hs_1/2,h^2s_2/4\}\). For every
\(0<\eps\le2\eta\), there exist \(c^+,c^-\in\A\) with
\(\|c^+-c^-\|_\infty=\eps\) and
\(\bigl|(D_2c^+)_{i_\star}-(D_2c^-)_{i_\star}\bigr|=4\eps/h^2\).
If \(c^0\) is globally strictly feasible and \(0<\eps<2\eta\), then
\(c^+\) and \(c^-\) are globally strictly feasible.
\end{theorem}

\begin{proof}
Set \(\alpha=\eps/2\). Define \(r_{i_\star-1}=\alpha\),
\(r_{i_\star}=-\alpha\), \(r_{i_\star+1}=\alpha\), and \(r_j=0\)
otherwise. Let \(c^\pm=c^0\pm r\). Then
\(\|c^+-c^-\|_\infty=2\|r\|_\infty=\eps\).

On the affected rows, listed in increasing index order,
\[
(D_1r)_{I_1}=\frac{\alpha}{h}(-1,2,-2,1),\qquad
(D_2r)_{I_2}=\frac{\alpha}{h^2}(1,-3,4,-3,1).
\]
Since \(\eps\le2\eta\), \(\|r\|_\infty=\eps/2\le\kappa\),
\(\|D_1r\|_\infty=\eps/h\le s_1\), and
\(\|D_2r\|_\infty=2\eps/h^2\le s_2\). Thus the affected price-bound, two-sided \(D_1\), and \(D_2\) constraints remain satisfied; all other
constraints are unchanged, so \(c^+,c^-\in\A\). Finally, \((D_2r)_{i_\star}=4\alpha/h^2\), so
\((D_2c^+)_{i_\star}-(D_2c^-)_{i_\star}=4\eps/h^2\). If \(c^0\) is
globally strictly feasible and \(\eps<2\eta\), all affected inequalities are
strict and all other constraints retain their original strictness.
\end{proof}

Thus no-arbitrage feasibility alone does not eliminate the sharp local
amplification (Figure~\ref{fig:mechanism-overview}).

\subsection{Curvature in Risk Outputs}

Let \(c(\cdot)\) denote a twice differentiable normalized call-price curve
whose values on the grid form the vector \(c\). Under
\(V(S,K)=Sc(x)\), where \(V\) is USD option value, \(S\) is spot, \(K\) is
strike, \(F\) is the forward, \(q:=F/S\), and \(x:=K/F=K/(qS)\), holding
\(K\), \(q\), and \(c(\cdot)\) fixed,
\[
\Delta_S:=\frac{\partial V}{\partial S}=c(x)-x c'(x),\qquad
\Gamma_S:=\frac{\partial^2 V}{\partial S^2}=\frac{x^2}{S}c''(x).
\]
The curve-induced grid quantity \(x_i^2(D_2c)_i/S\) inherits
the bound in~\eqref{eq:universal-d2-bound}, scaled by \(x_i^2/S\).
The empirical analysis replaces \(D_2\) by its nonuniform counterpart \(D_2^x\).
The empirical analysis evaluates reported Greeks with the Black formula under
the inverse-option convention; these are separate functionals, and no algebraic
equality with this curve-induced quantity is assumed.

\subsection{Bounds on Local Curvature Repair Risk}

\begin{definition}[Local curvature repair risk]
For nonempty \(\A\) and \(i=1,\ldots,N-1\), define the local curvature
repair risk by
\[
\Risk_{D_2}(i;\A)
:=
\max_{c\in\A}(D_2c)_i-\min_{c\in\A}(D_2c)_i.
\]
Because \(\A\) is compact and the objective is linear, both extrema are
attained and can be computed by linear programs.
\end{definition}

\begin{theorem}[Bounds on local curvature repair risk]
\label{thm:binding-safe}
Assume \(\A\neq\varnothing\). Fix \(i_\star\in\{3,\ldots,N-3\}\). If some
\(c^0\in\A\) satisfies Theorem~\ref{thm:sharp-admissible-amplification}'s
local slack assumptions, then
\[
\begin{aligned}
\Risk_{D_2}(i_\star;\A)&\ge\frac{8\eta}{h^2}
=\min\left\{\frac{8\kappa}{h^2},\frac{4s_1}{h},2s_2\right\},\\
\eta&:=\min\{\kappa,hs_1/2,h^2s_2/4\}.
\end{aligned}
\]
Independently of this slack assumption, for any interior \(i\) and
\(\delta>0\), if \(a_j-b_j\le\delta\) for \(j=i-1,i,i+1\), then
\(\Risk_{D_2}(i;\A)\le\min\{1/h,4\delta/h^2\}\).
\end{theorem}

\begin{proof}
The lower bound follows from
Theorem~\ref{thm:sharp-admissible-amplification} with \(\eps=2\eta\).
For the upper bound, any \(c,\tilde c\in\A\) satisfy
\(|c_j-\tilde c_j|\le\delta\) on the stencil, giving
\(4\delta/h^2\). Moreover,
\((D_2c)_i=((D_1c)_{i-1}-(D_1c)_i)/h\), so the call-spread and convexity
constraints imply \(0\le(D_2c)_i\le1/h\) for every \(c\in\A\). Hence the
feasible range is at most \(1/h\). Combining these bounds proves the result.
\end{proof}

\begin{remark}[Interpretation of the lower bound]
For fixed \(\kappa,s_1,s_2\), the lower bound
\(\min\{8\kappa/h^2,4s_1/h,2s_2\}\) is nonincreasing in \(h\): denser grids
raise the price-bound and call-spread terms until the convexity term binds.
\end{remark}

\section{Data and Validation Design}

The theory highlights grid spacing \(h\), box slack \(\kappa\), the minimum slack
\(s_1\) to either call-spread bound, convexity slack \(s_2\), and a local upper
bound \(\delta\) on effective price-interval widths. Rather than estimating
the local slacks \(\kappa,s_1,s_2\) separately, we use the linear-programming
(LP) feasible range, which captures the joint effect of the effective price
bounds and the imposed no-arbitrage constraints. The remainder of this section
describes the data, feasible set, repair rules, and matched design used to
evaluate repair risk empirically.

\paragraph{Data and normalization.}
The empirical study uses hourly Deribit BTC and ETH inverse (coin-settled) option chains represented in normalized call-equivalent prices; linear USDC-settled options are excluded. The main 12:00 UTC sample contains 375,894 quotes with timestamps, maturities, strikes, forwards, spots, prices, quality flags, and relative spreads. Under the inverse-option convention, the call premium \(c\), quoted in units of the base currency, satisfies \(c=\Phi_{\rm N}(d_1)-(K/F)\Phi_{\rm N}(d_2)\), with intrinsic value \(\max\{1-K/F,0\}\). Here, \(K\) is strike, \(F\) is the recorded Deribit forward, \(\Phi_{\rm N}\) is the standard normal CDF, and \(d_1,d_2\) are the usual Black quantities based on forward moneyness. Calls enter the repair grid directly. Let \(c_i\) and \(p_i\) denote normalized call and put premiums at strike \(K_i\). Puts are mapped to the same normalized call-price coordinate by put--call parity, \(c_i=p_i+1-K_i/F\), with bid and ask endpoints transformed consistently. The repair grid contains one call-equivalent observation per distinct strike within each maturity slice and is ordered by strike. After repair, put prices and put Greeks used in book templates are recovered by the same parity relation. The recorded forward \(F\) defines the common normalized-strike and price
scale before repair and is held fixed across all matched repair comparisons.

Let \(x_i:=K_i/F\) be normalized strike. Empirically, we set \(h_i=(x_{i+1}-x_{i-1})/2\) for interior observations, while endpoints use adjacent spacing. With \(h_-=x_i-x_{i-1}\) and \(h_+=x_{i+1}-x_i\),
\[
(D_2^x c)_i=\frac{2}{h_-+h_+}
\left\{\frac{c_{i+1}-c_i}{h_+}-\frac{c_i-c_{i-1}}{h_-}\right\}.
\]
The absolute row-sum norm is \(4/(h_-h_+)\), reducing to \(4/h^2\) on a uniform grid, so the corresponding nonuniform amplification scale replaces \(h^2\) by \(h_-h_+\). For interior observations, \(h_i=(h_-+h_+)/2\), so \(h_i^{-2}\le (h_-h_+)^{-1}\). We use \(\mathrm{RelSpread}_i/h_i^2\), written \(\mathrm{RelSpread}/h^2\), as a symmetrized proxy for the nonuniform amplification scale when ranking slices. Here \(\mathrm{RelSpread}_i=(\mathrm{ask}_i-\mathrm{bid}_i)/\max\{|y_i|,5\times10^{-4}\}\), with \(y_i\) the normalized mid price.

We also restrict repair evaluations by maturity: \(2<\mathrm{DTE}\le30\), where DTE denotes days to expiry. Marking and next-day book measures use \(0<\mathrm{DTE}\le31\) with one extra marking day.

\paragraph{Evaluation units and slice selection.}
We use nested aggregation so inference is run at the calendar-month cluster
level rather than the strike level. A slice is one option chain for a fixed
currency, date, and maturity, observed at 12:00 UTC. Evaluations pool selected
slices by currency, month, and selection group; currency-month cells aggregate
selection groups within a currency and month; and calendar-month clusters
aggregate BTC and ETH. For the VaR-style analysis, cells are additionally
indexed by book template.

The classical design compares the smoothness path with the matched
\(L_1\)--\(L_2\) control in 108 evaluations: BTC and ETH over 18 sampled months,
2022-11 through 2026-05, each crossed with three
slice-selection groups: high-score, random, and low-score. Only slices with a nonempty within-maturity admissible set (\(\A\)) enter the ranking. Slices are ranked by their 90th-percentile
\(\mathrm{RelSpread}/h^2\) score. The high-score and low-score selections take
50 slices from the top and bottom of this ranking, while the random selection
draws 50 slices without conditioning on the score. The quote-spacing screen provides a simple threshold baseline for the ranking score.
From the same evaluation pool, it selects wing quotes with short times to
expiry when both relative spread and local spacing in normalized strike exceed
their respective 75th percentiles within the relevant currency and DTE bucket.
The \(\mathrm{RelSpread}/h^2\) screen selects the same pool's top score decile.
Every pair of repair rules is evaluated on the same selected slices, so
selection changes where repair risk is measured while keeping the input data
fixed across rules. For temporal coverage, the design begins with six anchor months (2022-11, 2023-03, 2023-06,
2024-03, 2025-01, and 2026-05). We then add 12 months sequentially, each chosen to maximize its minimum
distance from the months already selected, with ties resolved in favor of the
earlier month. Evaluation months and timestamps are fixed before repair outcomes are computed. The resulting design contains 36 currency-month cells and 18
calendar-month clusters.
The targeted \NNProj{} check asks whether its learned target yields a repair
distinct from classical repairs. Its 10 matched-slice evaluations span 6
currency-month cells across 4 months:
2023-03 selection comparisons, 2023-06 high-score evaluations, and high-score
out-of-sample windows.

\paragraph{Feasible set and repair rules.}
Feasibility is imposed separately at each maturity. Repaired prices must
remain within the intersection of the quoted bid--ask intervals and the
model-free bounds \([\max\{1-x_i,0\},1]\), satisfy
\(0\le(c_i-c_{i+1})/(x_{i+1}-x_i)\le1\), and satisfy
\((D_2^x c)_i\ge0\) at every interior strike. In the empirical analysis, \(\A\) denotes the within-maturity feasible set
defined by these nonuniform-grid constraints. Let \(y=(y_i)\) denote the vector
of normalized midpoint prices, \(\epsilon_p>0\) the price-width floor, and
\(c^{\rm NNProj,raw}\) the unconstrained neural price vector. Set
\(W_1=\mathrm{diag}(\max\{\mathrm{ask}_j-\mathrm{bid}_j,\epsilon_p\}^{-1})\).
Let \(W_2\) be diagonal, with squared weights
\(\max\{\mathrm{ask}_j-\mathrm{bid}_j,\epsilon_p\}^{-2}\) clipped within each
slice at its 5th and 95th percentiles. The four repair rules below represent
weighted projection, curvature regularization, and projection toward a learned
target within the same feasible set. For \(\rho\ge0\), they are defined by
\[
\begin{aligned}
\hat c^{L_1}&\in\arg\min_{c\in\A}\|W_1(c-y)\|_1,\\
\hat c^{L_2}&=\arg\min_{c\in\A}\|W_2(c-y)\|_2^2,\\
\hat c^{\rho}&=\arg\min_{c\in\A}\|W_2(c-y)\|_2^2+\rho\|D_2^x c\|_2^2,\\
\hat c^{\rm NNProj}&=\arg\min_{c\in\A}\|W_2(c-c^{\rm NNProj,raw})\|_2^2 .
\end{aligned}
\]
We denote \(\hat c^\rho\) by \(\mathrm{Smooth}_\rho\), with \(\rho\in\{0.01,0.1,1,10\}\). The \(\rho\)-grid defines the regularization path; \(\rho\) has no economic interpretation independent of the strike grid. Reported experiments use \(\epsilon_p=5\times10^{-4}\) and the numerical floor \(\epsilon_{\rm num}=10^{-12}\). Since the \(L_1\) minimizer may not be unique, the deterministic HiGHS
optimizer \cite{huangfuHall2018highs} under fixed solver settings and tolerances defines the
implemented \(L_1\) repair. Using linear interpolation over the overlapping
normalized-strike range, a two-maturity calendar screen tests persistence under
\(c_{\rm long}(x_g)\ge c_{\rm short}(x_g)\).

\paragraph{Learned and parametric target repair checks.}
\NNProj{} is a learned repair: a neural fit to total implied variance followed
by projection. Its multilayer perceptron (MLP) fits \(\log w\) on past samples,
where \(\sigma_{\rm imp}\) is implied volatility,
\(\mathcal T\) is year-fraction time to expiry, and
\(w:=\sigma_{\rm imp}^2\mathcal T\) is total implied variance, using features
\(k=\log(K/F)\), \(|k|\), \(k^2\),
\(k^3\), \(\log\mathcal T\), \(\log(1+\mathrm{DTE})\), and
\(\log(1+\mathrm{RelSpread})\). The network standardizes features and uses
hidden layers \((64,64)\), tanh activations, Adam, an \(\ell_2\) weight penalty of \(10^{-4}\),
at most 500 iterations, early stopping, and a 15\% validation split. Predicted
total implied variance is converted to \(c^{\rm NNProj,raw}\) and projected back
onto \(\A\). The targeted \NNProj{} check uses up to 12 months of history; the
seed replication tests sensitivity to training seeds after at least 12 observed
training months. Both separate
training and evaluation with a 7-day embargo. The reported coefficient of
determination \(R^2\) uses quote rows from the evaluation window.

Separate from the classical study, the seed replication uses slices with the
highest \(\mathrm{RelSpread}/h^2\) scores in
18 months, each with a full 12-month lookback. Before examining seed outcomes,
we excluded the 2025-01 and 2026-05 design pilots and selected the months by
maximin spacing, retaining the first and last eligible months and resolving
ties in favor of the earlier month. This design yields 36 currency-month cells
and 180 refits. The \(D_2^x\) comparisons use slices on which all five
projections succeed. Across refits, the training window, architecture, selected slices, feasible set,
and projection operator are fixed; the replication measures randomized training
variation. Finally, the
stochastic-volatility-inspired (SVI) target check asks whether a standard
parametric target also selects a distinct admissible repair. Its raw target
\(c^{\rm SVI,raw}\) is obtained from a five-parameter raw SVI fit to midpoint
total implied variance \cite{gatheral2014arbitrage}. The SVI repair uses the
smoothness objective above at \(\rho=0.01\), with
\(y\) replaced by \(c^{\rm SVI,raw}\), leaving \(\A\) unchanged.

\begin{figure}[htp]
\centering
\includegraphics[width=0.92\columnwidth]{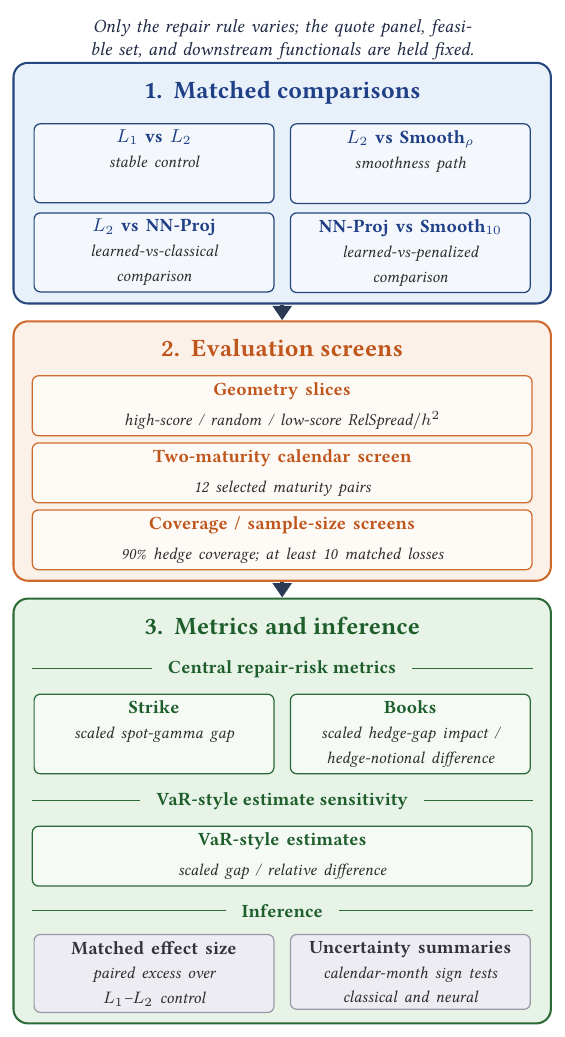}
\Description{Single-column flowchart of the matched validation protocol, showing matched comparisons, evaluation screens, metrics, and inference.}
\caption{Matched validation protocol. The protocol fixes the quote panel,
feasible set, and downstream functionals, then compares repair rules
through evaluation screens, repair-risk metrics, and inference.}
\label{fig:validation-protocol}
\end{figure}

\paragraph{Matched validation protocol.}
Figure~\ref{fig:validation-protocol} summarizes the matched protocol; the
formal metrics are defined below.

{\sloppy
\paragraph{Metrics.}
\emph{Strike.} We compare repairs by the absolute spot-gamma difference scaled
by the pairwise median absolute gamma within each slice. Let \(\Gamma_i^{(m)}\) denote spot gamma at
strike \(i\) under rule \(m\).  Operationally, each repaired normalized call price is inverted to implied volatility and evaluated with the same Black marking formula under the inverse-option convention to obtain USD spot delta and gamma. Let \(\mathcal S(i)\) denote the evaluation slice containing strike observation \(i\); all repair rules in a pair share this slice. For repair rules \(m,n\) and slice label \(\lambda=\mathcal S(i)\), \(\Delta^\Gamma_{mn,i}=|\Gamma_i^{(m)}-\Gamma_i^{(n)}|/M^\Gamma_{mn,\lambda}\), where \(M^\Gamma_{mn,\lambda}=\max\{\operatorname{median}_{j:\mathcal S(j)=\lambda,\ r\in\{m,n\}}|\Gamma_j^{(r)}|,\epsilon_{\rm num}\}\). The median is over strikes \(j\) in slice \(\lambda\) and the two repair rules in the matched comparison. We report 90th percentiles (p90) within each evaluation.\looseness=-1

\begin{figure*}[htp]
\centering
\includegraphics[width=0.90\textwidth]{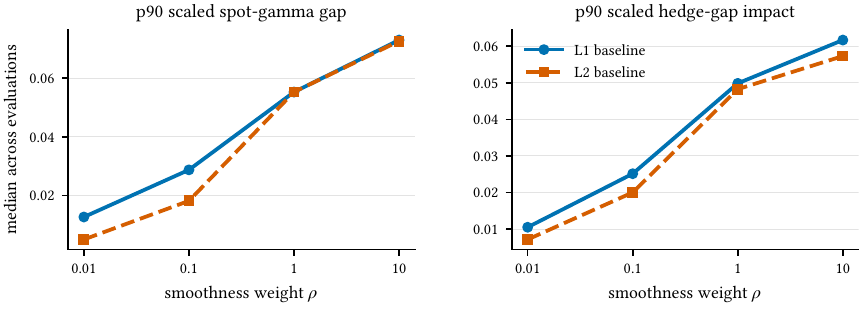}
\Description{Baseline to smoothness rho path for L_1 and L_2 baselines across four smoothness weights.}
\caption{Classical smoothness path. Baseline-to-smoothness dispersion is nondecreasing in \(\rho\) under both the \(L_1\) and \(L_2\) baselines.}
\label{fig:rho-path}
\end{figure*}

Separately, the feasible-set curvature range uses the nonuniform operator \(D_2^x\): \(\Risk_{D_2^x}(i;\A)=\max_{c\in\A}(D_2^x c)_i-\min_{c\in\A}(D_2^x c)_i\), computed by two LPs for each selected interior strike. For the LP-range comparison, an observation is counted as an exceedance when its observed absolute
\(D_2^x\) gap exceeds
\(\Risk_{D_2^x}(i;\A)(1+10^{-5})+10^{-10}\). In Results, tables, and figures, empirical \(D_2^x\) quantities are abbreviated as \(D_2\).

\emph{Books and VaR-style estimates.} The treatment of missing next-day
quotes can affect sample composition and measured book losses~\cite{zhongWu2026markingAware}. We therefore require 90\% next-day
coverage for evaluation-level hedge and book summaries. VaR-style
cells require at least 10 matched next-day losses.
Let \(C_k,P_k\) denote nearest available call/put legs at target
\(k=\log(K/F)\). Standardized templates use fixed gross spot notional: an ATM straddle
\(C_0+P_0\), a put spread \(P_{-0.05}-P_{-0.10}\), a risk reversal
\(C_{0.05}-P_{-0.05}\), and a put butterfly
\(P_{-0.10}-2P_{-0.05}+P_0\). These templates target level, wing-slope, skew,
and wing-convexity exposures, respectively.

For a fixed book and evaluation cell at date \(t\), let \(B_t^{(m)}\) be the repaired book value and \(H_t^{(m)}\) the hedge position, defined as minus the book delta computed from repaired USD spot deltas and fixed leg weights. Let \(S_t\) denote spot at date \(t\). The primary hedge metric is scaled hedge-gap impact: the absolute hedge gap times the absolute next-day spot move, divided by the largest absolute value among the common and repaired book valuations for the two rules, with the denominator floored at USD~1. Separately, hedge notional gaps are
\(10^4|H_t^{(m)}-H_t^{(n)}|/\sum_\ell|\nu_\ell|\) bps and are summarized by
their median p90; \(S_t\sum_\ell|\nu_\ell|\) is gross spot notional and
\(\nu_\ell\) are fixed leg quantities.

Write \(\Delta\Pi_{t+1}^{(m)}\) for next-day hedged book change, including book revaluation, hedge P\&L, and transaction cost. The common-scenario calculation fixes the book legs, observed \(t\) and
\(t+1\) option marks, spot move, and denominator across repairs, so only the
hedge varies. When the book is revalued using each repair, that repair determines
the current and next-day book values, hedge, and denominator. The corresponding
normalized loss is \(-\Delta\Pi_{t+1}^{(m)}/\max\{|B_t^{(m)}|,1\}\), with
transaction cost set to zero in reported experiments. Together, the two
calculations distinguish hedge changes from \mbox{full revaluation}.

VaR-style estimates are empirical 95\% quantiles of next-day normalized losses
matched across repairs within each book-template and selection-group cell. Because these cells contain 10--50 next-day observations, the estimates are descriptive measures of tail loss rather than regulatory backtests. Let \(u\) index a valid cell, identified by currency month, book template, and selection group, and let \(v_{m,u}\) be its VaR-style estimate under repair rule \(m\). The pairwise VaR-style gap is scaled as
\(|v_{m,u}-v_{n,u}|/\max\{|v_{m,u}|,|v_{n,u}|,\epsilon_{\rm num}\}\).
For comparisons with \(L_2\), we also report the absolute relative difference
\[
\omega_{m,u}:=
\frac{|v_{m,u}-v_{\mathrm{L}_2,u}|}
{\max\{|v_{\mathrm{L}_2,u}|,\epsilon_{\rm num}\}}.
\]

\emph{Inference.} For matched comparisons, we compute paired excesses over the matched \(L_1\)--\(L_2\) control. Since selection groups within a currency month are correlated, we first aggregate paired excesses within currency-month cells. For classical smoothness, we then take the median across BTC and ETH cells within each calendar month before the sign test. For the neural seed replication, the primary excess is the median pairwise p90 \(D_2\) gap on slices where all five projections succeed minus the matched \(L_1\)--\(L_2\) control; BTC and ETH are aggregated by calendar-month median. We report two-sided exact sign tests at the calendar-month level and, for classical smoothness, the corresponding one-sided test.
\par}

\section{Results}

\subsection{Smoothness Selects Distinct Admissible Repairs}

\paragraph{Main result.}

Table~\ref{tab:evidence-dashboard} collects the headline measures. Since \(L_1\) and \(L_2\) use different weighting rules, their small
spot-gamma gap provides the matched projection-rule control. Across the 108 classical control evaluations, the median p90 scaled spot-gamma gap is 0.0039 for \(L_1\) versus \(L_2\).

\begin{table*}[htp]
\centering
\small
\setlength{\tabcolsep}{2.2pt}
\renewcommand{\arraystretch}{0.96}
\caption{Summary of main empirical results.}
\label{tab:evidence-dashboard}
\begin{tabularx}{\textwidth}{@{}>{\raggedright\arraybackslash}p{0.16\textwidth}>{\raggedright\arraybackslash}p{0.33\textwidth}>{\raggedright\arraybackslash}p{0.25\textwidth}Y@{}}
\toprule
Claim / comparison & Measure and sample & Estimate & Diagnostic \\
\midrule
\(L_1\) vs.\ \(L_2\) control & Median evaluation-level p90 scaled spot-gamma gap; 108 evaluations & 0.0039 & Small control gap \\
\(L_2\) vs.\ \(\mathrm{Smooth}_{10}\) & Same measure; 108 evaluations and 18 calendar-month clusters & Gap 0.0726; excess 0.0686 & Median month excess 0.0651; 18/18 positive; two-sided \(p=7.63\times10^{-6}\) \\
LP feasible \(D_2\) range & Observed \(D_2\) gap versus LP range; 51{,}189 high-score interior strikes & Spearman 0.8722; log--log \(R^2=0.6112\) & LP-bound exceedances: 57/51{,}189 initially; 0 after tighter-tolerance recomputation \\
Targeted \NNProj{} check & Spot-gamma gaps on successful projections; 10 evaluations, 500/500 problems & \(L_2\)--\NNProj{} 0.1890; \NNProj{}--\(\mathrm{Smooth}_{10}\) 0.1957; control 0.0108 & Positive in 6/6 currency-month cells and all 4 calendar months \\
Five-seed replication & Pairwise \(D_2\) gaps; 36 high-score currency-month cells and 18 calendar months & 1.3822 versus control 0.0719; paired excess 1.2003 & 17/18 months positive; two-sided \(p=1.45\times10^{-4}\); slices shared across seeds: median 100\%, minimum 80.9\% \\
Downstream books / VaR-style estimates & Hedge: 42 risk-reversal evaluations; VaR-style: 432 common-scenario and 417 repair-specific cells & Hedge 18.62 bps; VaR-style 0.0111/0.0933 and 0.0904/0.3882 & Full-revaluation gaps are descriptively larger \\
\bottomrule
\end{tabularx}
\vspace{1pt}
\begin{minipage}{0.99\textwidth}
\footnotesize
\textit{Notes:} Spot gamma entries are median p90s across evaluations; VaR
entries are median/p90 over valid cells: common scenario first, full
revaluation under each repair second.
\end{minipage}
\end{table*}

Along the implemented \(\rho\)-path, smoothness regularization changes only
the selection criterion within the fixed feasible set (Figure~\ref{fig:rho-path}). The median p90 scaled spot-gamma gap for
\(L_1\) versus \(\mathrm{Smooth}_\rho\) rises from 0.0126 at \(\rho=0.01\) to
0.0730 at \(\rho=10\). For \(L_2\) versus \(\mathrm{Smooth}_\rho\), it rises
from 0.0050 to 0.0726. The scaled hedge-gap impact follows the same pattern: its median p90 rises from 0.0105 to 0.0617 for \(L_1\) and from 0.0071 to 0.0573 for \(L_2\).

The paths indicate a systematic smoothness effect: spot-gamma gaps are nondecreasing in 108/108 evaluations under both baselines; hedge-gap impacts are nondecreasing in 95/108 under \(L_1\) and 96/108 under \(L_2\). At \(\rho=10\), the evaluation-level excess of \(L_2\)--\(\mathrm{Smooth}_{10}\) over the \(L_1\)--\(L_2\) control is 0.0686. After currency-month and calendar-month aggregation, the
excess remains positive in all 18 months, with median 0.0651 and exact sign
test \(p=3.81\times10^{-6}\) one-sided (\(7.63\times10^{-6}\) two-sided).

\paragraph{Robustness checks.}
The median calendar-month excess is similar across the six anchor months and 12 added
months (0.0678 versus 0.0639). The two-maturity calendar screen on 12 selected
maturity pairs leaves the observed dispersion essentially unchanged: the median
post-screen/pre-screen dispersion ratio is 1.000. An 18-month SVI-target check
extends the result to a standard parametric smile target:
\(L_2\) versus the SVI-target repair is 0.0834, and its excess over the matched
\(L_1\)--\(L_2\) control is positive in 108/108 evaluations. The SVI-target
projection succeeds on 5{,}376/5{,}400 preselected problems.

\subsection{Locating and Bounding Repair Risk}

To locate candidate slices, we compare two screens across 108 evaluations and
classical repair pairs: median p90 scaled spot-gamma gaps are 0.0457 on 103
nonempty quote-spacing screens and 0.0651 on 95 nonempty
\(\mathrm{RelSpread}/h^2\) screens. Within each evaluation, regressing \(\log(1+\Delta^\Gamma)\) on log relative
spread and \(\log(1/h_i^2)\) gives both coefficients positive in 100/108 evaluations
but weak fit (median \(R^2=0.0290\)), so we use \(\mathrm{RelSpread}/h^2\) only as a slice-ranking score.

\begin{figure}[htp]
\centering
\includegraphics[width=\columnwidth]{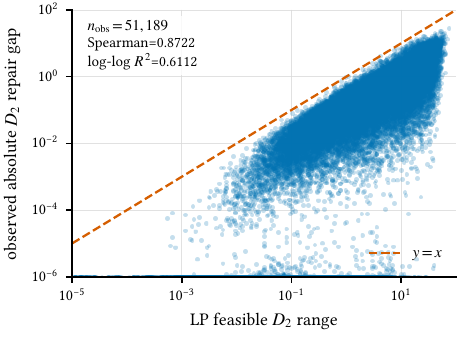}
\Description{Log-log scatter plot comparing the LP feasible D2 range with the observed absolute D2 repair gap. Nearly all points lie below the identity line.}
\caption{Feasible curvature range and observed repair gap. The plot compares
the LP feasible \(D_2\) range with the observed \(D_2\) gap for
\(L_2\)--\(\mathrm{Smooth}_{10}\) high-score slices; the dashed line is the
identity line, and values below \(10^{-6}\) are shown at the plotting floor.}
\label{fig:lp-range-scatter}
\end{figure}

We then compute the feasible curvature range from \(\A\): two LPs per selected interior strike give the \(D_2\)-range. In high-score \(L_2\)--\(\mathrm{Smooth}_{10}\) slices across 36 currency-month cells and 18 calendar months, this range has Spearman correlation 0.8722 with observed \(D_2\) curvature gaps and log--log \(R^2=0.6112\). Under the stated numerical criterion, 57/51{,}189 observations exceed the computed LP range. Recomputing these 57 cases with tighter solver tolerances yields no exceedance at an absolute tolerance of \(10^{-9}\). Figure~\ref{fig:lp-range-scatter} supports the feasible-set mechanism: the score locates slices, while the LP range measures the curvature variation allowed by \(\A\) and bounds realized gaps.

\subsection{Learned Repairs and Sensitivity to Training Seeds}

The \NNProj{} results show that predictive fit and no-arbitrage feasibility are
distinct requirements. Across 10 evaluations, training samples contain 10,719
to 45,412 option quotes, and held-out \(R^2\) for log total implied variance
has median 0.941, with range 0.858 to 0.973. Despite this predictive performance, before
projection, \NNProj{} prices fall outside the bid--ask intervals for a median
55.9\% of quotes, with a median price mean absolute error (MAE) of 2.21
bid--ask widths. Prediction is evaluated in log total implied variance, whereas
feasibility is imposed in normalized price units, so projection into \(\A\) is
required. Slice eligibility is fixed before repair, and all 500 preselected
\NNProj{} surfaces project successfully within solver tolerance; subsequent
comparisons therefore concern admissible repairs, while next-day book coverage
is assessed separately.

\begin{figure}[htp]
\centering
\includegraphics[width=0.94\columnwidth]{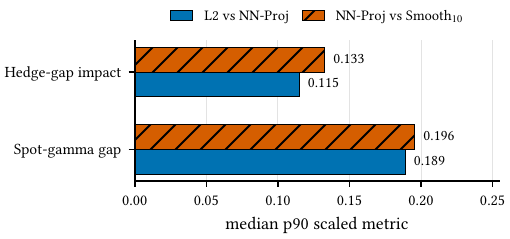}
\Description{Pairwise spot-gamma and hedge-gap measures comparing the learned repair with the L_2 and Smoothness-10 repair rules.}
\caption{Targeted learned-repair check. Post-projection \NNProj{} is compared with \(L_2\) and \(\mathrm{Smooth}_{10}\) in spot-gamma and book hedge measures.}
\label{fig:nnproj}
\end{figure}

\paragraph{Learned repair versus classical repairs.}
On matched slices with successful projections, sharing the same feasible set
does not make \NNProj{} coincide with \(L_2\) or
\(\mathrm{Smooth}_{10}\). Across the 10 evaluations, the median 90th
percentile of the scaled spot-gamma gap is 0.1890 for \(L_2\) versus \NNProj{} and
0.1957 for \NNProj{} versus \(\mathrm{Smooth}_{10}\), compared with 0.0108 for the matched
\(L_1\)--\(L_2\) control. Currency-month paired excesses are positive in 6/6 cells for both
comparisons and remain positive in 4/4 calendar months using the same median
rule. Median p90 scaled hedge-gap impact is 0.1149
for \(L_2\) versus \NNProj{} and 0.1328 for \NNProj{} versus
\(\mathrm{Smooth}_{10}\). Figure~\ref{fig:nnproj} shows that both the
spot-gamma gap and hedge-gap impact are larger for \NNProj{} versus
\(\mathrm{Smooth}_{10}\) than for \NNProj{} versus \(L_2\).

\begin{samepage}
\paragraph{Seed sensitivity.}
In the 18-month seed replication, the median p90 nonuniform \(D_2\) gap across seed pairs on
slices where all five projections succeed is 1.3822, versus 0.0719 for the
matched \(L_1\)--\(L_2\) control, with
median paired excess 1.2003. The excess is positive in 34/36 currency-month
cells and 17/18 months after BTC/ETH aggregation (two-sided exact
\(p=1.45\times10^{-4}\)). These common slices comprise a median 100\% and a
minimum 80.9\% of selected slices across windows. Despite similar price fit
after projection, curvature gaps remain large: the worst refit's
root-mean-square error (RMSE), scaled by bid--ask width, exceeds the best by at
most 8.17\% across windows (median 1.10\%).
\par
\end{samepage}

\begin{figure}[!b]
\centering
\includegraphics[width=0.88\columnwidth]{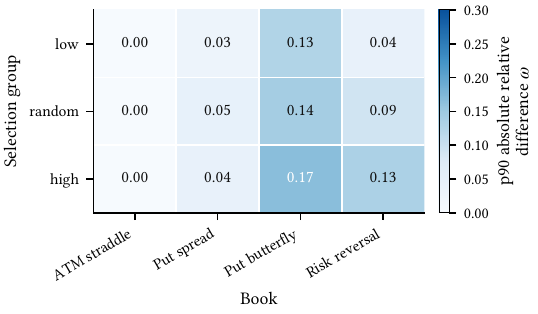}
\Description{Absolute relative differences in common-scenario VaR-style estimates by standardized book and selection group for Smoothness-10 relative to L2.}
\caption{Absolute relative differences in common-scenario VaR-style estimates.
Cells report the 90th percentile of the absolute relative difference
\(\omega_{m,u}\) for \(\mathrm{Smooth}_{10}\) relative to \(L_2\).}
\label{fig:var-relative-difference}
\end{figure}

\subsection{Downstream Hedge and VaR-Style Effects}

We first report standardized-book results across all classical repair
comparisons, followed by separate results for \(L_2\)--\(\mathrm{Smooth}_{10}\)
and \(L_2\)--\NNProj{}. Figure~\ref{fig:var-relative-difference} reports
common-scenario absolute relative differences in VaR-style estimates for
\(\mathrm{Smooth}_{10}\) relative to \(L_2\).
The largest hedge differences occur in risk reversals,
whose call and put legs lie on opposite wings. Across 171 standardized book cells that pass the coverage screen, median next-day pair coverage is 0.9787 and the median p90 scaled hedge-gap impact is 0.0080. ATM straddles have a median p90 impact of 0.0002, while risk reversals reach 0.0330 because repairs affect their two legs differently.

Across the 42 risk-reversal evaluations passing the coverage screen,
the median p90 absolute hedge-notional gap between \(L_2\) and
\(\mathrm{Smooth}_{10}\) is 18.62 bps, equivalent to USD~1,862
of hedge position per USD~1 million gross spot notional. Across valid cells for the four standardized books, the scaled pairwise
VaR-style gap has median/p90 0.0111/0.0933 under the common-scenario
calculation (432 cells) and 0.0904/0.3882 when the book is revalued
using each repair (417 cells).

Under the absolute relative difference \(\omega_{m,u}\), the put butterfly
book shows the largest common-scenario VaR-style sensitivity, unlike the
hedge-gap ranking above. For the matched \(L_2\)--\NNProj{} comparison, the 90th percentile is 0.1226 and the
median is 0.0597. The absolute denominator \(|v_{\mathrm{L}_2,u}|\) is floored at
\(10^{-12}\); no reported common-scenario cell triggers this floor.
Under repair-specific revaluation, the corresponding 90th-percentile
differences are 0.7400 for \(L_2\)--\(\mathrm{Smooth}_{10}\) and 0.6100 for
\(L_2\)--\NNProj{} among their respective valid cells. These values are larger than when only the hedge varies, but the sets of valid
cells differ.

\section{Conclusion and Implications}
Option-surface price feasibility is an input to model validation, not a
substitute for it. Remaining within bid--ask intervals and satisfying static no-arbitrage across strikes do
not ensure stable downstream Greeks, hedge quantities, or risk measures. The
matched design shows that smoothness penalties and projection toward a learned target change downstream risk quantities while
holding the raw quote panel, feasible set, and downstream functionals
fixed. Theorem~\ref{thm:binding-safe} gives the upper bound
\(\min\{1/h,4\delta/h^2\}\) on local curvature repair risk, where \(\delta\)
bounds the three effective interval widths and \(h\) is normalized-strike
spacing. Together with the locally sharp construction in
Theorem~\ref{thm:sharp-admissible-amplification}, this shows that narrow
effective price intervals need not imply tight curvature identification on
dense strike grids. The upper bound holds for any pair of admissible repairs,
including smoothness-penalized repairs, learned repairs, and seed refits of
the same learned model, while the quote panel and feasible set stay fixed.
The classical comparisons cover 18 sampled months from November 2022
to May 2026, with additional evidence from neural seed refits and a
two-maturity calendar screen.

The implication extends beyond the sampled BTC and ETH windows:
when quote bounds and no-arbitrage constraints admit multiple repairs,
the repair rule is part of the risk model. Validation should therefore
assess dispersion in Greeks and hedge quantities across matched repairs,
alongside price fit and no-arbitrage feasibility. The LP curvature range
quantifies the local variation permitted by the full admissible set and
helps identify slices that warrant closer review. Future work can examine
how repair choice affects stochastic volatility calibration and subsequent
pricing and hedging.

\interlinepenalty=10000
\bibliographystyle{ACM-Reference-Format}
\bibliography{references}

\end{document}